\documentclass{aastex}          
\usepackage{spr-astr-addons}    

\begin{document}

\title{A transition from a decelerated to an accelerated phase of the universe expansion from the simplest non-trivial polynomial function of $T$ in the $f(R,T)$ formalism}

\shorttitle{<Short article title>}
\shortauthors{}

\author{P.H.R.S. Moraes\altaffilmark{1}}
\author{G. Ribeiro\altaffilmark{2}}
\author{R.A.C. Correa\altaffilmark{3}}

\affil{$^{1}$ITA - Instituto Tecnol\'ogico de Aeron\'autica - Departamento de F\'isica, 12228-900, S\~ao Jos\'e dos Campos, S\~ao Paulo, Brazil}
\affil{$^{2}$FEG - Faculdade de Engenharia de Guaratinguet\'a - Departamento de F\'isica e Qu\'imica, 12516-410, Guaratinguet\'a, S\~ao Paulo, Brazil}
\affil{$^{3}$CCNH, Universidade Federal do ABC, 09210-580, Santo Andr\'{e}, SP,
Brazil}
\email{moraes.phrs@gmail.com} 
\email{gabozo_ribeiro@hotmail.com} 
\email{rafael.couceiro@ufabc.edu.br}

\begin{abstract}
In this work we present cosmological solutions from the simplest non-trivial $T$-dependence in $f(R,T)$ theory of gravity, with $R$ and $T$ standing for the Ricci scalar and trace of the energy-momentum tensor, respectively. Although such an approach yields a highly non-linear differential equation for the scale factor, we show that it is possible to obtain analytical solutions for the cosmological parameters. For some values of the free parameters, the model is able to predict a transition from a decelerated to an accelerated expansion of the universe.
\end{abstract}

\keywords{cosmological models - cosmic acceleration - $f(R,T)$ gravity}

\section{Introduction}\label{sec:int}

With the modern observational astrophysics, it was discovered that the universe expansion is accelerating. If one considers the universe to be composed mostly by matter, this acceleration is highly non-intuitive since gravity, as an attractive force, should slow down the expansion velocity. The observation of Type Ia Supernovae (\cite{riess/1998,perlmutter/1999}) seems to support a universe made mostly ($\sim70\%$) by an exotic component dubbed dark energy (DE). The DE has an equation of state (EoS) $\omega\sim-1$ which justifies the apparent anti-gravitational aspect of the present universe dynamics. In $\Lambda$CDM cosmology, the DE is mathematically described by a cosmological constant (CC) ($\Lambda$) in the Einstein's field equations (FEs). However, the CC, coincidence and dark matter problems, missing satellites, hierarchy problem and other shortcomings (see (\cite{clifton/2012}) and references therein) arising from $\Lambda$CDM model yield the consideration of alternative cosmological models.

We are led, then, to search for some kind of matter fields which could generate the observable outcomes without the $\Lambda$CDM shortcomings. For instance, scalar fields which slowly go down to their potential can produce sufficient negative pressure in order to make the universe expansion to accelerate (\cite{caldwell/1998,tsujikawa/2013,moraes/2014,khurshudyan/2014,khurshudyan/2015,farooq/2011}). 

Other efficient alternatives to find answers for the cosmological issues mentioned above come from the family of $f(R)$ (\cite{defelice/2010}), (\cite{sotiriou/2010}) and $f(R,T)$ (\cite{harko/2011}) theories; the latter focusing not only in generalizing the geometrical terms - those proportional to the Ricci scalar $R$ - in the gravitational part of the action, but also its matter content - proportional to the trace of the energy-momentum tensor $T$. 

Those alternative gravity theories have already originated well-behaved cosmological scenarios, as one can check in (\cite{sepehri/2016,farajollahi/2012,ms/2016,mc/2016,rao/2015,sahoo/2015,shamir/2015,singh/2015,moraes/2015,moraes/2015b,moraes/2014b,ahmed/2014}).

The cosmological features of the above models are directly related to the functional forms of the $f(R)$ or $f(R,T)$ functions. 
The simplest non-trivial polynomial function of the Ricci scalar $R$ in the $f(R)$ theories was used in a quite seminal work (\cite{starobinsky/2007}). A.A. Starobinsky has proposed a class of models for the dependence of the gravitational part of the action on $R$. Those have reproduced $\Lambda$CDM features at recent times and satisfied solar system and laboratory tests. Today, the most popular of these models, known as Starobinsky model (SM) is the one for which the functional form of $f(R)$ is given by $R+\alpha R^2$, with $\alpha$ being a constant. 

Our purpose in the present article is to propose a Starobinsky-like model for the dependence on $T$ in the $f(R,T)$ formalism, i.e., we will derive a cosmological scenario from $f(R,T)=f(R)+f(T)$, with $f(T)=\alpha T+\beta T^{2}$, and $\alpha$ and $\beta$ being constants. This is the simplest non-trivial polynomial function of $T$ for the functionality of $f(R,T)$ as is the SM in $f(R)$ gravity and due to the high non-linearity of the resulting differential equations it has not been used so far in $f(R,T)$ gravity for cosmological or any other purposes.

An $f(R)=R+\alpha R^{2}$ gravity can generate matter bounce cosmological solutions, as shown in (\cite{oikonomou/2015}). Observational constraints on the SM parameter can be found in (\cite{dev/2008}). Furthermore, the structure of neutron stars in SM was discussed in (\cite{ganguly/2014,orellana/2013,thongkool/2009}) and the collapse of massive stars in (\cite{goswami/2014}). Studies of gravitational waves in SM can be appreciated in (\cite{yang/2011,capozziello/2009}).

The cosmological features of a Starobinsky-like $T$-dependence in $f(R,T)$ gravity will be presented below. In Section \ref{sec:frt} a brief review of $f(R,T)$ gravity is exposed. We present solutions for the correspondent cosmological parameters, such as scale factor, Hubble parameter and deceleration parameter in Section \ref{sec:rab}. The model is constructed by assuming a matter-dominated universe, i.e., without necessity of assuming the universe dynamics is dominated by a CC or some sort of quintessence. In Section \ref{sec:dis} we discuss our results.

\section{A brief review of the $f(R,T)$ gravity}\label{sec:frt}

The gravitational part of the action in the $f(R,T)$ gravity is given by

\begin{equation}\label{frt1}
S_G=\frac{1}{16\pi}\int f(R,T)\sqrt{-g}d^4x,
\end{equation}
with $f(R,T)$ being the general function of the Ricci scalar $R$ and the trace of the energy-momentum tensor $T$, and $g$ the determinant of the metric. According to the authors, such a $T$-dependence comes from the consideration of quantum effects which are neglected in $f(R)$ gravity, for instance. Moreover, throughout this work, we will consider units such that the gravitational constant and the speed of light are equal to $1$.

The FEs of the $f(R,T)$ gravity are obtained by varying (\ref{frt1}) with respect to the metric and read:

\begin{align}\label{frt2}
&f_{R}(R,T)R_{\mu\nu}-\frac{1}{2}f(R,T)g_{\mu\nu}+(g_{\mu\nu}\Box-\nabla_{\mu} \nabla_{\nu})\nonumber \\
&f_{R}(R,T)=8\pi T_{\mu\nu}-f_{T}(R,T)T_{\mu\nu}-f_{T}(R,T)\Theta_{\mu\nu},
\end{align}
in which $\Theta_{\mu\nu}=-2T_{\mu\nu}-pg_{\mu\nu}$, $R_{\mu\nu}$ is the Ricci tensor, $T_{\mu\nu}=diag(\rho,-p,-p,-p)$ is the energy-momentum tensor, which we are assuming to be the one of a perfect fluid, with $\rho$ and $p$ representing the matter-energy density and pressure of the universe, respectively, $\Box$ is the D'Alambert operator and $\nabla$ is the covariant derivative. Moreover, $f_{R}(R,T)$ and $f_{T}(R,T)$ are the partial derivatives of $f(R,T)$ with respect to $R$ and $T$, respectively.

\section{The $f(R,T)=R+\alpha T$+$\beta T^{2}$ cosmology}\label{sec:rab}

We are going to assume for $f(R,T)$ the simplest non-trivial polynomial function of $T$ and the simplest dependence on $R$, i.e., $f(R,T)=R+\alpha T+\beta T^{2}$ in (\ref{frt1}), with $\alpha$ and $\beta$ being constants, and $T=\rho-3p$. Such a functional form benefits from the fact that one can recover General Relativity (GR) just by letting $\alpha=\beta=0$. 

Recall that \cite{harko/2011} have proposed a generalization of $f(R)$ theories, by making the gravitational part of the action to depend generally not only on its geometrical terms, but also on its matter terms. We are dealing here with the same GR geometrical term on the action, but we are generalizing the matter terms (those proportional to $T$) in order to check if such a generalization is able to generate well-behaved cosmological scenarios in the same way SM does in $f(R)$ gravity (\cite{starobinsky/2007,borowiec/2012,odintsov/2015}).

By substituting $f(R,T)=R+\alpha T+\beta T^{2}$ in Eq.(\ref{frt2}) yields:

\begin{align}\label{frt4}
G_{\mu\nu}=&8\pi T_{\mu\nu}+\alpha\left[T_{\mu\nu}+\frac{1}{2}(\rho-p)g_{\mu\nu}\right]+\nonumber \\
&2\beta(\rho-3p)\left[T_{\mu\nu}+\frac{1}{4}(\rho+p)g_{\mu\nu}\right],
\end{align}
for which $G_{\mu\nu}$ stands for the usual Einstein tensor. We have written Eq.(\ref{frt4}) in such an elegant form in order to explicit the terms proportional to $\alpha$ and $\beta$. We can see that the terms proportional to $\alpha$ carry linear corrections in the matter terms while the terms proportional to $\beta$ carry quadratic corrections. 

By developing Eq.(\ref{frt4}) for a Friedmann-Robertson-Walker metric with null curvature, we obtain 

\begin{equation}\label{frt5}
3\left(\frac{\dot{a}}{a}\right)^{2}=8\pi\rho+\frac{1}{2}[\alpha(3\rho-p)+\beta(5\rho+p)(\rho-3p)],
\end{equation}
\begin{equation}\label{frt6}
2\frac{\ddot{a}}{a}+\left(\frac{\dot{a}}{a}\right)^{2}=-8\pi p+\frac{1}{2}[\alpha (\rho-3p)+\beta (\rho-3p)^{2}],
\end{equation}
with $a$ being the scale factor and dots representing time derivatives.

Let us consider a matter EoS for the universe, i.e., $p=0$. Such an assumption yields the following differential equation for the scale factor:

\begin{equation}
\frac{\ddot{a}}{a}+\frac{3}{10}\left(  \frac{\dot{a}}{a}\right)  ^{2}+\left(
\frac{1}{2}-\frac{4}{3}\pi\sqrt{\frac{6}{5\beta}}\right)  \frac{\dot{a}}%
{a}=0,\label{frt7}
\end{equation}
where we are assuming that $\alpha=-16\pi/3$.

We can see that (\ref{frt7}) has a non-linear character. It might be important to remark that the presence of nonlinearity is not surprising, given that such a behavior is found in a wide range of areas of Physics nowadays (\cite{sdc/2009,sdc/2010,csd/2015,cmr/2015,cmsdr/2015}). Because of the nonlinearity, we are led to ask if the problem can be analytically resolved.

We show below that, indeed, it is possible to obtain an analytical solution for the scale factor differential equation above. To do this, let us firstly multiply Eq.(\ref{frt7}) by the factor $(a/\dot{a})$, obtaining:

\begin{equation}
\frac{\ddot{a}}{a}+\frac{3}{10}\frac{\dot{a}}{a}+\frac{1}{2}-\frac{4}
{3}\pi\sqrt{\frac{6}{5\beta}}=0.\label{frt8}
\end{equation}

Now, after straightforward manipulations, it is easy to deduce from Eq.(\ref{frt8}) that the equation for the scale factor can be put in the form
\begin{equation}
\frac{d}{dt}\left(\ln\dot{a}+\lambda_{1}\ln a\right)  =-\lambda
_{2},\label{frt9}
\end{equation}
where $\lambda_{1}\equiv3/10$ and $\lambda_{2}\equiv1/2-4\pi\sqrt{6/(5\beta)}/3$.

Therefore, by solving Eq.(\ref{frt9}), we obtain
\begin{equation}
a(t)=a_{0}\left(e^{-\lambda_{2}t}+c_{0}\right)^\frac{10}{13},\label{frt11}
\end{equation}
where $a_{0}$ and $c_{0}$ are arbitrary constants of integration. 

Using Equation (\ref{frt11}) we are able to find the Hubble parameter $H=\dot{a}/a$ and the deceleration parameter $q=-\ddot{a}/(\dot{a}H)$, for which negative values of the latter indicate an accelerated expansion of the universe. 

From (\ref{frt11}) we derive

\begin{equation}\label{frt12}
H=-\frac{10}{13}\frac{\lambda_2}{c_0e^{\lambda_2t}+1},
\end{equation}
\begin{equation}\label{frt13}
q=-\left(\frac{13}{10}c_0e^{\lambda_2t}+1\right).
\end{equation}

Below we are going to depict the evolution of $H$ and $q$ through time. Presumably, different values of $\beta$ imply different behaviors for the cosmological parameters. Moreover, by invoking the initial condition $a(0)=0$, we obtain $c_0=-1$.

\begin{figure}[ht!]
\vspace{0.3cm}
\centering
\includegraphics[height=5cm,angle=00]{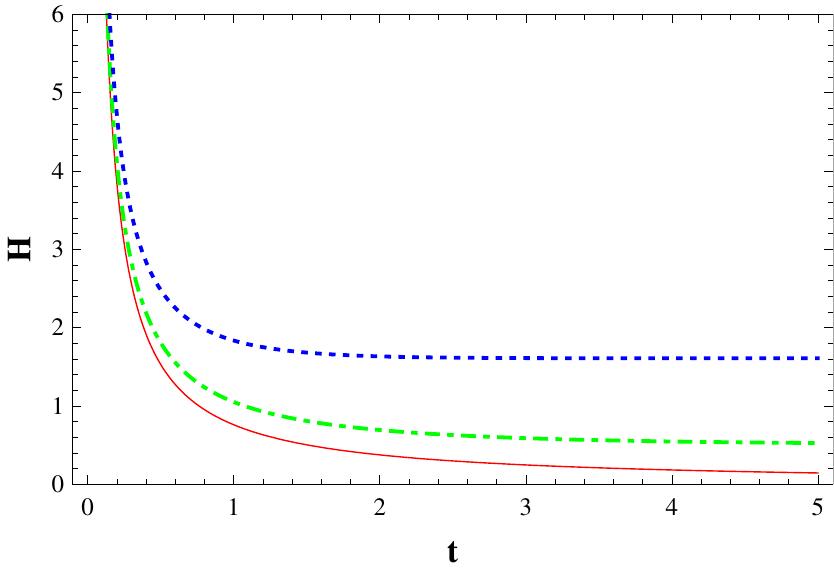}
\caption{Time evolution of the Hubble parameter from Equation (\ref{frt12}). The (blue) dotted line stands for $\beta=\pi$, while the (green) dot-dashed and (red) solid lines for $\beta=5\pi$ and $\beta=30\pi$, respectively.}
\label{fig1}
\end{figure}  

\begin{figure}[ht!]
\vspace{0.3cm}
\centering
\includegraphics[height=5cm,angle=00]{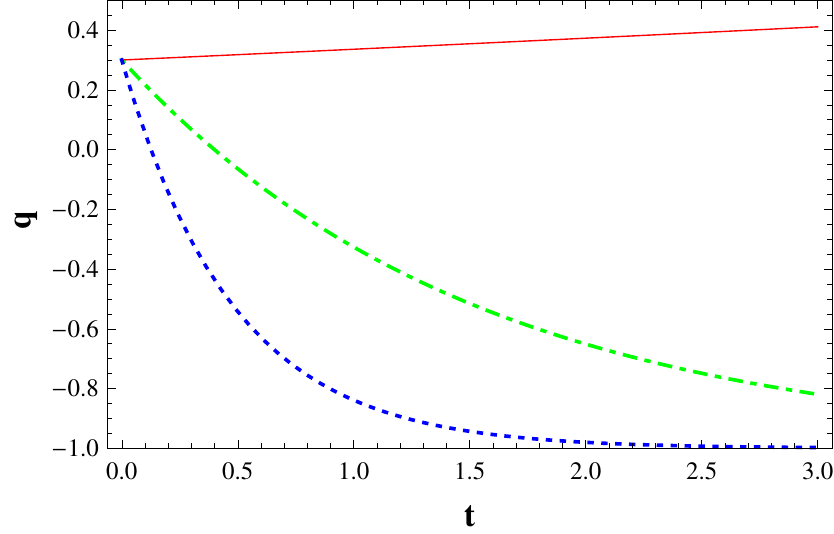}
\caption{Time evolution of the deceleration parameter from Equation (\ref{frt13}). The (blue) dotted line stands for $\beta=\pi$, while the (green) dot-dashed and (red) solid lines for $\beta=5\pi$ and $\beta=30\pi$, respectively.}
\label{fig2}
\end{figure}

In the next section we are going to interpret the behavior of the curves in Figs.\ref{fig1}-\ref{fig2} above.

\section{Discussion}\label{sec:dis}

The CC problem is one of the greatest mysteries of Cosmology today. Although $\Lambda$CDM model can provide a great matching between theoretical predictions and observations, it still lacks a convincing explanation for the DE physical interpretation. There is a huge discrepancy between the observed value of the CC density (\cite{hinshaw/2013}) and its value obtained from Particle Physics theoretical predictions (\cite{weinberg/1989}). Since the CC is considered to be the responsible for the cosmic acceleration, it is worth trying to describe such a dynamical phenomenon without invoking it.

Rather than invoking the CC and consequently the shortcomings related to its physical interpretation, $f(R,T)$ theories present in their FEs extra terms arising from a general dependence on $R$ and/or $T$. Those extra terms may be responsible for the cosmic acceleration, as we will argue below.

Departing from what can be seen today in the literature, we have proposed an $f(R,T)$ functionality whose geometrical dependence resembles the GR one, while it contains in its matter content, linear and quadratic correction terms, as $f(T)=\alpha T+\beta T^{2}$. As quoted above, this can be seen as a Starobinsky-like dependence for $T$, since the so-called SM in $f(R)$ gravity is such that $f(R)=R+\alpha R^{2}$. Our proposal in this article was to check if such an assumption can, in the same way SM does, generate well-behaved cosmological models, which can predict the cosmic acceleration. Let us check this in the next paragraphs.

Eq.(\ref{frt7}) is the equation one obtains when considering the Starobinsky-like model for the $T$-dependence and a standard $R$-dependence in $f(R,T)$ gravity. In such an equation we have taken $p=0$, i.e., we are assuming the universe dynamics is dominated by matter. Moreover we have taken $\alpha=-16\pi/3$ for the sake of its construction.

We have presented Eq.(\ref{frt11}) as the Eq.(\ref{frt7}) solution. From it, we were able to derive Eqs.(\ref{frt12})-(\ref{frt13}) as the Hubble and deceleration parameters, respectively.

The time evolution of these cosmological parameters is depicted in Figs.\ref{fig1}-\ref{fig2}. We have shown the behaviors of $H$ and $q$ for different values of $\beta$. The constant $c_0$ was set to $-1$ because of the initial condition $a(0)=0$. Moreover, when plotting $H$ and $q$ there was no necessity of assuming any value for $a_0$ in (\ref{frt11}).

In Fig.\ref{fig1} we can see that for different values of $\beta$, the predicted Hubble parameters are well-behaved. Firstly they are all restricted to positive values, which is expected in an expanding universe. Also, in standard cosmology the Hubble parameter is proportional to the inverse of the Hubble time $t_H$, as $H\propto1/t_H$. Such a feature is also being respected in Fig.\ref{fig1}.

Fig.\ref{fig2} depicts the evolution of the deceleration parameter for different values of $\beta$. The deceleration parameter is defined as $-\ddot{a}a/\dot{a}^{2}$, so that negative values of it describe an acceleration of the universe expansion. We can see that for $\beta=\pi$ and $\beta=5\pi$ the universe expansion has previously slowed down its velocity and then reached an epoch in which it speeded up. Such a scenario still prevails and is known as DE era in standard cosmology. On the other hand, from the (red) solid line curve in Fig.\ref{fig2}, we realize that for $\beta=30\pi$ it is not possible to describe an accelerating expansion, since $q$ is restricted to increasing positive values.

Let us analyse such features from the perspective of the $\lambda_2$ parameter values. We can see that for $\beta<128\pi^{2}/15$, $\lambda_2<0$. From Eq.(\ref{frt13}), we see that in order to be able to predict an accelerating universe ($q<0$ after a certain period of time), $\lambda_2$ must be negative. On the contrary, $q$ will always assume increasing positive values (recall that $c_0=-1$). 

The constraint $\beta<128\pi^{2}/15$ shall be respected in future works with the same functionality of $f(R,T)$ since an accelerating expansion is an indispensable cosmic feature today.

To finish, we would like to remark that the material correction terms proportional to linear and quadratic functions of $\rho$ and $p$ are, indeed, the responsible for the cosmic acceleration in the present model. Consider Eq.(\ref{frt6}) for a matter-dominated universe with $\alpha=\beta=0$. The solution for the scale factor in this case is $a(t)=c_1(3t-2c_2)^{2/3}$. Such a scale factor yields $q=1/2$ independently of the values of the constants $c_1$ and $c_2$. Apart from the fact that such a deceleration parameter is constant, it is positive, i.e., it is not in agreement with an accelerated expansion. Therefore, we have shown that linear and quadratic material corrections in the gravity action can generate a well-behaved cosmological model, with a varying deceleration parameter which through its evolution predicts a transition from a decelerated to an accelerated phase of the universe expansion.

\

{\bf Acknowledgements} 

PHRSM would like to thank S\~ao Paulo Research Foundation (FAPESP), grant 2015/08476-0, for financial support. RACC thanks to UFABC and CAPES for financial support.

\bibliographystyle{spr-mp-nameyear-cnd}  

\end{document}